# Quantum Saturation of the Electro-Optic Effect

Aiden Ross[1,*], Sankalpa Hazra[1], Albert Suceava[1], Dylan Sotir[2,3], Darrell G. Schlom[2,4,5], Venkatraman Gopalan[1], and Long-Qing Chen[1]

Future quantum computing architectures require electro-optic materials that maintain a strong, stable performance at cryogenic temperatures. In conventional electro-optic materials, large electro-optic coefficients are often confined to narrow temperature windows near structural phase transitions, where small changes in temperature lead to large changes in the electro-optic response. Using thermodynamic analysis, phase-field simulations, experimental growth and cryogenic optical measurements we show that quantum fluctuations can be harnessed to overcome this trade-off. By tuning the ferroelectric phase boundaries down to 0 K, quantum fluctuations induce a saturation regime in which a large electro-optic response becomes nearly temperature-independent below 25 K. We demonstrate that the phase boundaries can be tuned through either strain in $BaTiO_3$ or through chemical composition in $Ba_{1-x}Ca_xTiO_3$, leading to a large, temperature insensitive, cryogenic electro-optic effect comparable to bulk $BaTiO_3$ at room temperature; the performance exceeds $BaTiO_3$-on-Si by over an order of magnitude. These findings establish a general design principle for engineering high-performance electro-optic materials for cryogenic applications.

[1]Department of Materials Science and Engineering and Materials Research Institute, The Pennsylvania State University, University Park, PA 16802

[2]Department of Materials Science and Engineering, Cornell University, Ithaca, New York 14853, USA

[3]Platform for the Accelerated Realization, Analysis, and Discovery of Interface Materials (PARADIM), Cornell University, Ithaca, New York 14853, USA

[4]Kavli Institute at Cornell for Nanoscale Science, Ithaca, New York 14853, USA

[5]Leibniz-Institut für Kristallzüchtung, Max-Born-Straße 2, 12489 Berlin, Germany

* Email Address: amr8057@psu.edu

# Introduction

Electro-optic materials play a central role in emerging quantum photonic technologies. As practical quantum computers scale to millions of qubits, they will rely on a modular architecture linking many quantum processors with low-loss and low noise photonic links[1,2], made possible by high-performance electro-optic materials that read[3,4], write[3,5], and transfer quantum states[6]. These same materials underpin photonic quantum computing platforms[7]. Across all these potential applications a large cryogenic electro-optic effect reduces operating voltages, increases efficiency, and minimizes device footprint.

$BaTiO_3$ has emerged as a promising material for electro-optic devices, with a large electro-optic coefficient (bulk: $r_{51}$ ~1500 pm/V,[8] film: $r_{51}$ ~900 pm/V)[9] at room temperature. Unfortunately its performance sharply degrades at cryogenic temperatures ($r_{eff}$ ~200 pm/V)[9,10]. Theoretical calculations attribute its large room-temperature electro-optic coefficient to the close competition between these different ferroelectric states in the proximity to the tetragonal-to-orthorhombic ferroelectric phase transition[11], leading to a dynamic averaging of lower symmetry monoclinic structures[12]. Nonetheless, this room temperature response is both highly temperature dependent and strongly suppressed at cryogenic temperatures due to phase transitions[10,11,13]. A central goal is therefore to identify strategies that enhance the electro-optic response of $BaTiO_3$ while simultaneously stabilizing it against variations in temperature.

At low temperatures, quantum fluctuations dominate the behaviour of ferroelectrics, giving rise to quantum saturation effects[14]. In quantum paraelectrics, such as $SrTiO_3$ and $KTaO_3$, quantum fluctuations suppress the ferroelectric phase altogether, despite an underlying soft mode instability[15–17]. In $BaTiO_3$, by contrast, quantum fluctuations do not eliminate the ferroelectric phase, but instead suppress changes in the ferroelectric polarization below a characteristic saturation temperature $(T_s)$[18]. In this quantum-saturated regime, quantum fluctuations dominate over thermal fluctuations, leading to properties which become temperature insensitive[14], presenting a compelling opportunity to realize a large, thermally stable electro-optic response. So far realizing materials in which a strong electro-optic effect coincides with the quantum saturation regime has remained a challenge. Recent theoretical calculations and experimental measurements have demonstrated the possibility of using epitaxial strain to substantially enhance the cryogenic electro-optic effect in $BaTiO_3$ thin films[19]. Nevertheless, the relationship between the enhanced properties and quantum fluctuations remains largely unexplored.

Although epitaxial strain provides a powerful tuning knob to enhance the electro-optic response, strain relaxation limits the film thickness[20,21] thereby constraining the degree of optical confinement for practical electro-optic devices[22]. These limitations motivate the development of alternative design strategies that create a large quantum saturated electro-optic response without relying upon epitaxial strain.

In this work, we show how the conventional trade-off between the magnitude and thermal stability of the electro-optic response can be overcome by quantum fluctuations. Using phase-field simulations coupled with a thermodynamic analysis of optical properties we first establish a direct relationship between the temperature-strain phase diagram and the electro-optic response in BaTiO$_3$. Quantum fluctuations of the ferroelectric soft mode are incorporated using a Barrett-type modification to the Landau coefficients[23], leading to a quantum saturated electro-optic response over a moderate temperature window. Experimental measurements probe the saturation effect in BaTiO$_3$ thin films grown on GdScO$_3$ substrates, which are in quantitative agreement with the phase-field simulations and exceed the performance of previous BaTiO$_3$ thin films by a factor of 14[13] and recent work on isotope-exchanged SrTiO$_3$ by a factor of 2.5[24]. We further extend this design strategy to compositionally tuned Ba$_{1-x}$Ca$_x$TiO$_3$, which shows that a large quantum saturated electro-optic response can be achieved without epitaxial strain. Together, these results establish a general design strategy to engineer high-performance electro-optic materials for quantum photonic technologies.

## Results

### Strain Tuning of the Electro-Optic Effect

**Figure 1a** presents the temperature-strain phase diagram of BaTiO$_3$ thin films. The phase at each point is largely determined by a combination of two thermodynamic driving forces: the intrinsic chemical stability of each ferroelectric phase, quantified by the Landau energy, and the elastic energy contribution, which favours phases closely matching the imposed strain. For a BaTiO$_3$ crystal in the absence of any applied stress/strain or electric field, the Landau energy determines the ranges of temperatures for the most stable ferroelectric state: the tetragonal phase (*P*4*mm*) from 400 K to 270 K, the orthorhombic phase (*Amm*2) from 270 K to 200 K and the rhombohedral phase (*R*3*m*) below 200 K[25]. The elastic energy, in contrast, biases the system toward phases or domain structures that best accommodate the imposed strain. A biaxial compressive strain promotes the ferroelectric phases or domain configurations that possess an out-of-plane component of the spontaneous polarization, such as the tetragonal *c*-phase. As the compressive strain increases, the tetragonal *c*-phase becomes increasingly favoured, and beyond -0.8%, the 'as grown' phase diagram suggests that all ferroelectric–ferroelectric transitions are eliminated, leaving only the tetragonal *c*-phase (**Figure 1a**).

Strain can be leveraged to promote phase competition, creating regions of the phase diagram where multiple ferroelectric phases are nearly thermodynamically degenerate, resulting in a close energetic competition. In some cases, instead of forming a two-phase mixture, the system can resolve this phase competition by adopting an intermediate, lower-symmetry phase, that bridges between these two competing phases[26,27]. Phase-field simulations reveal two distinct monoclinic phases (**Figure 1a**). The first is a monoclinic M$_a$ phase (space group *Cm*), which acts as a bridge between the tetragonal and rhombohedral phases, marked by the orange region of the phase diagram, and the second is a monoclinic M$_c$ phase (space group *Pm*), which acts as a bridge between the tetragonal and orthorhombic phases, marked by the pink region of the phase diagram. **Figure 1b** and **1c** show the simulated $r_{51}$ and $r_{11}$ electro-optic response as a function of temperature and strain. Both the $r_{51}$ and $r_{11}$ electro-optic coefficients are maximized within the monoclinic M$_c$ phase, with this enhancement becoming increasingly more pronounced as temperature decreases. While the monoclinic M$_c$ phase has been observed in bulk BaTiO$_3$ in the vicinity of the tetragonal-to-orthorhombic phase transition[28–30], it is a metastable phase in the bulk and only exists within a narrow temperature range. By using epitaxial strain, this monoclinic phase can be thermodynamically stabilized over a broader region of the temperature-strain phase diagram, thus extending the window where the electro-optic response is enhanced. At low temperatures, quantum fluctuations of the ferroelectric soft mode suppress the temperature dependence of the ferroelectric phase boundaries leading to the temperature independent electro-optic response[18].

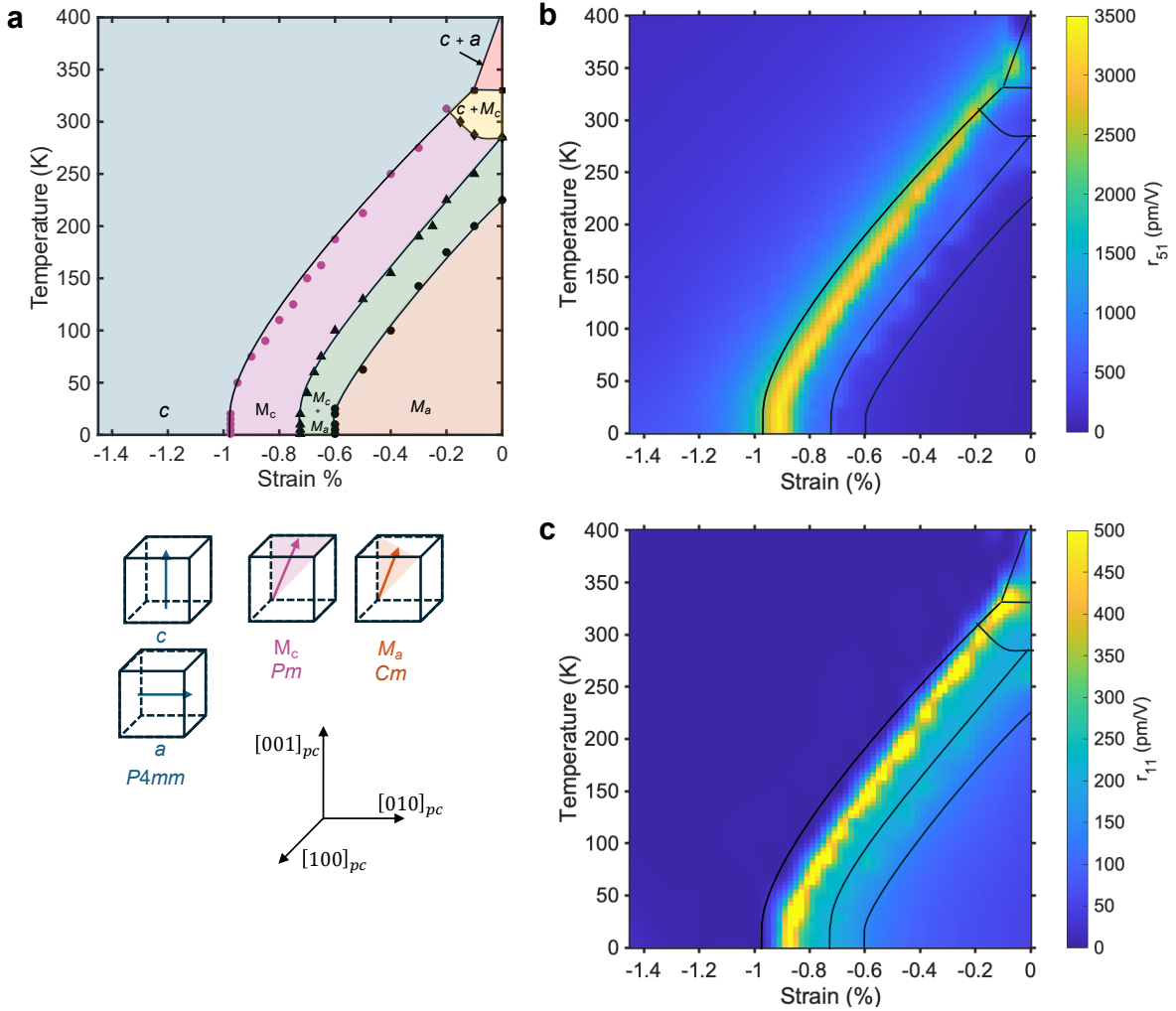

**Figure 1. Strain-engineered electro-optic response of BaTiO₃. a**, Temperature-strain phase diagram of BaTiO$_3$ showing the stability regions of the tetragonal (*c* out-of-plane, *a* in-plane), monoclinic M$_c$ and monoclinic M$_a$ phases. **b,c** Temperature and strain dependence of the linear electro-optic $r_{51}$ and $r_{11}$, respectively, for BaTiO$_3$, illustrating the relationship between the phase diagram and the electro-optic response.

**Figure 2** focuses on the strain-dependent properties at 1 K, highlighting how the spontaneous polarization, and the lattice dielectric response govern the electro-optic coefficients. **Figure 2a** shows the strain dependence of the $r_{11}$ and $r_{51}$ electro-optic coefficients. Notably, these two coefficients reach their maximum value at two distinct strain states, with $r_{11}$ peaking at -0.875% strain and $r_{51}$ peaking at -0.925% strain. This behaviour can be understood by examining the tensorial contributions to the electro-optic effect,

$$r_{ijk} \cong 2g_{ijmn} P_n^L \chi_{mk}^L \quad (1)$$

where $r_{ijk}$ is the electro-optic tensor, $g_{ijkl}$ is the quadratic polar-optic tensor, $P_i^L$ is the lattice polarization (the ionic and electronic contributions of the polarization that arises due to the displacement of the lattice) and $\chi_{ij}^L$ is the lattice dielectric susceptibility[11,31].

For BaTiO$_3$, which possesses a cubic ($m\bar{3}m$) parent phase, and assuming the off-diagonal components of the dielectric tensor remain small, the expressions for the $r_{11}$ ($r_{111}$) and $r_{51}$ ($r_{131}$) electro-optic coefficients simplify to

$$r_{111} \cong 2g_{1111} P_1^L \chi_{11}^L, \quad (2)$$
$$r_{131} \cong 2g_{1313} P_3^L \chi_{11}^L, \quad (3)$$

where $P_1^L$ and $P_3^L$ are the in-plane and out-of-plane components of the lattice polarization, respectively, and $\chi_{11}^L$ is the in-plane lattice dielectric susceptibility. From equation 2, the $r_{11}$ requires a non-zero $P_1^L$ and $r_{51}$ electro-optic coefficients require a non-zero $P_3^L$ (**Figure 2b**). The $\chi_{11}^L$ lattice dielectric constant is maximized in the M$_c$ phase at -0.925% strain, which coincides with the maximum for $r_{51}$ since $P_3^L$ remains non-zero across this region. Note that $r_{11}$ peaks at a different strain value since it is dependent upon $P_1^L$, which is nearly 0 at - 0.925% and increases with a decreasing strain. As a result, the product $P_1^L \chi_{11}^L$ is maximized at -0.875% rather than the strain where $\chi_{11}^L$ alone peaks, leading to distinct optimal conditions for $r_{11}$ and $r_{51}$.

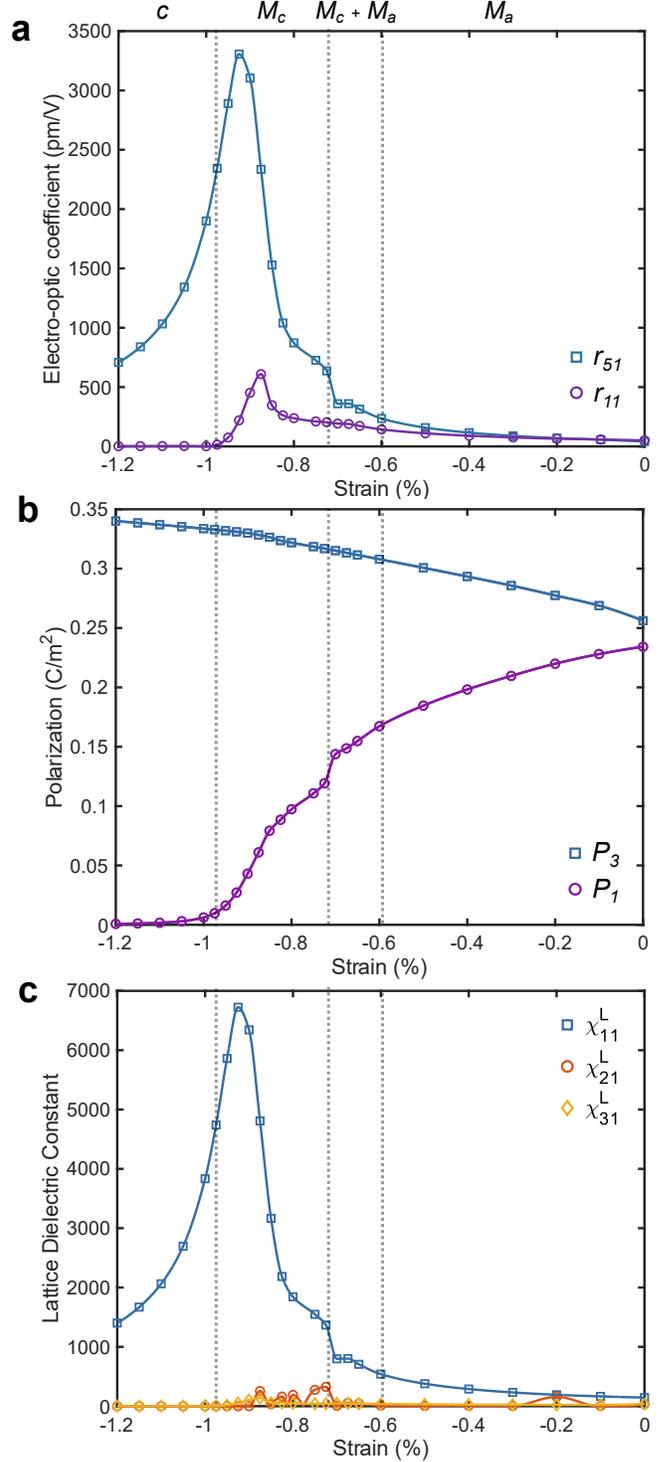

**Figure 2. Strain-dependent electro-optic response, polarization and dielectric properties at 0 K. a**, spatially averaged linear electro-optic coefficients ($r_{11}$ purple, $r_{51}$ blue). **b**, spontaneous polarization ($P_1$ purple, $P_3$ blue). **c**, the lattice dielectric constants ($\chi_{11}^L$ blue, $\chi_{21}^L$ orange, $\chi_{31}^L$ yellow) at 1 K. The dashed lines indicate the phase boundaries.

**Figure 3** further examines the temperature dependence of the $r_{51}$ and $r_{11}$ electro-optic coefficients under different strain conditions and demonstrates the quantum saturation behaviour. As shown in **Figure 3a**, the temperature at which $r_{51}$ reaches its maximum shifts to lower values as the compressive strain increases from 0% to -1.2% strain. This trend coincides with the strain-dependent onset of the M$_c$ phase reflected by the appearance of a non-zero $r_{11}$ electro-optic coefficient, which is forbidden by symmetry in the tetragonal phase, but is allowed when the polarization rotates away from the tetragonal axis (**Figure 3b**). Beyond -1% compressive strain, the peak of the $r_{51}$ coefficient shifts to 0 K and the magnitude of the $r_{51}$ coefficient decreases with further compressive strain as the system moves away from this phase-competition region.

Across all strain conditions, the electro-optic coefficient becomes nearly temperature independent as $T \to 0$ K reflecting the quantum saturation of the electro-optic response[14]. At room temperature, thermal fluctuations dominate and the corresponding electro-optic properties strongly vary with temperature. Importantly, below a characteristic saturation temperature ($T_s$) quantum fluctuations become comparable to, or larger than thermal fluctuations leading to a temperature independent electro-optic response[32]. Within a Landau mean-field treatment of the ferroelectric polarization, the fluctuations of the order parameter can be evaluated using either classical or quantum statistics[14,18]. When these fluctuation amplitudes are inserted into the mean-field Landau expansion, e.g.,

$$f_{Landau} = a_1(T)(P_1^2 + P_2^2 + P_3^2) + a_{11}(P_1^4 + P_2^4 + P_3^4) + a_{12}(P_1^2 P_2^2 + P_2^2 P_3^2 + P_2^2 P_3^2) + \cdots, \quad (4)$$

the classical result (dashed black line) produces the typical Curie-Weiss form of the $a_1(T)$ coefficient

$$a_1(T) = a_0(T - T_c), \quad (5)$$

as shown by the dashed line in **Figure 3c**. In contrast, the quantum treatment requires a modification to the classical Curie-Weiss form. Their effect can be captured by introducing a quantum temperature scale where $T^Q = T_s \coth\left(\frac{T_s}{T}\right)$,[32] where $T^Q \approx T$ at high temperature and $T^Q \approx T_s$ as $T \to 0$ K. Substituting $T^Q$ into the Curie-Weiss form yields a modified the Curie-Weiss law called the Barrett function[23],

$$a_1(T) = a_0(T^Q - T_c^Q) = a_0\left(T_s \coth\left(\frac{T_s}{T}\right) - T_s \coth\left(\frac{T_s}{T_c}\right)\right), \quad (6)$$

which follows the classical expression at high temperatures, but exhibit low temperature deviations due to quantum fluctuations (**Figure 3d**)[23]. When this quantum-corrected coefficient is used in the phase-field model, the calculated $r_{51}$ electro-optic coefficient exhibits a pronounced plateau (**Figure 3d solid line**), in stark contrast to the peak near 40 K expected from a purely classical model (**Figure 3d dashed line**).

To experimentally realize this quantum-saturated electro-optic effect, BaTiO$_3$ films are epitaxially grown on GdScO$_3$ (110)$_O$ substrates, where the subscript O denotes orthorhombic indices, which impose a biaxial compressive strain of approximately –1%. Phase-field simulations predict a substantial increase of the electro-optic response with a saturation regime below 30 K, which agree with experimental measurements of the effective electro-optic coefficient (**Figure 3e**. The presence of the monoclinic phase is further supported by the observation of a spontaneous polarization component along the [100]$_{pc}$ direction, which is supported using experimental second harmonic generation measurements (**Figure 3f**).

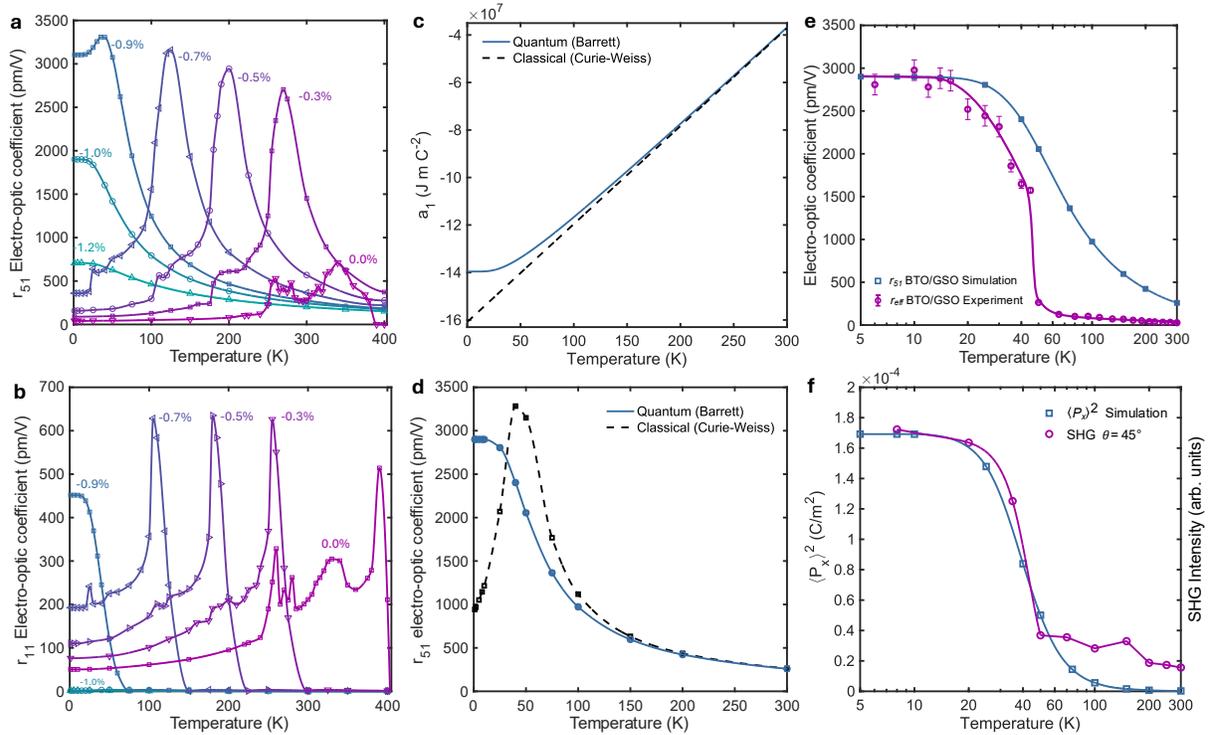

**Figure 3. Quantum saturation in the electro-optic effect.** Temperature dependence of the a, $r_{51}$ electro-optic coefficient and the b, the $r_{11}$ electro-optic coefficient. Increasing compressive strain shifts the maximum $r_{51}$ and $r_{11}$ to lower temperatures c, Temperature dependence of the a$_1$ Landau coefficient assuming classical and quantum fluctuations. d, Comparison of the $r_{51}$ electro-optic coefficient assuming classical and quantum fluctuations of BaTiO$_3$ epitaxially strained on GdScO$_3$. e Simulated and experimental electro-optic response of BaTiO$_3$ epitaxially strained on GdScO$_3$. f, Tetragonal-to-monoclinic phase transition probed by second harmonic generation (SHG)[15], compared with the simulated mean squared polarization along the [100]$_{pc}$ direction

## Chemical Tuning of the Electro-Optic Effect

While strain engineering is a powerful method for tuning the thermodynamics stability of different ferroelectric phases, it faces a fundamental limitation imposed by film thickness. In electro-optic devices, a key performance metric is the optical confinement factor, which quantifies the fraction of optical power guided within the electro-optic layer[22]. Thicker films increase the optical confinement, allowing for a larger fraction of the guided light to be actively modulated by the applied electric field. Unfortunately, the maximum film thickness is limited by the epitaxial strain, where beyond a critical thickness, strain relaxation occurs through the formation of dislocations[20,21]. While this relaxation can be mitigated using superlattice architectures, it nevertheless introduces an inherent trade-off between strain tuning and optical confinement.

Using the same design principle, one alternative strategy to mitigate this trade-off is to use composition as an additional tuning knob, offering a complementary pathway for optimizing cryogenic electro-optic materials. Importantly, compositional tuning does not impose a critical thickness enabling thicker electro-optic films with a large optical confinement factor.

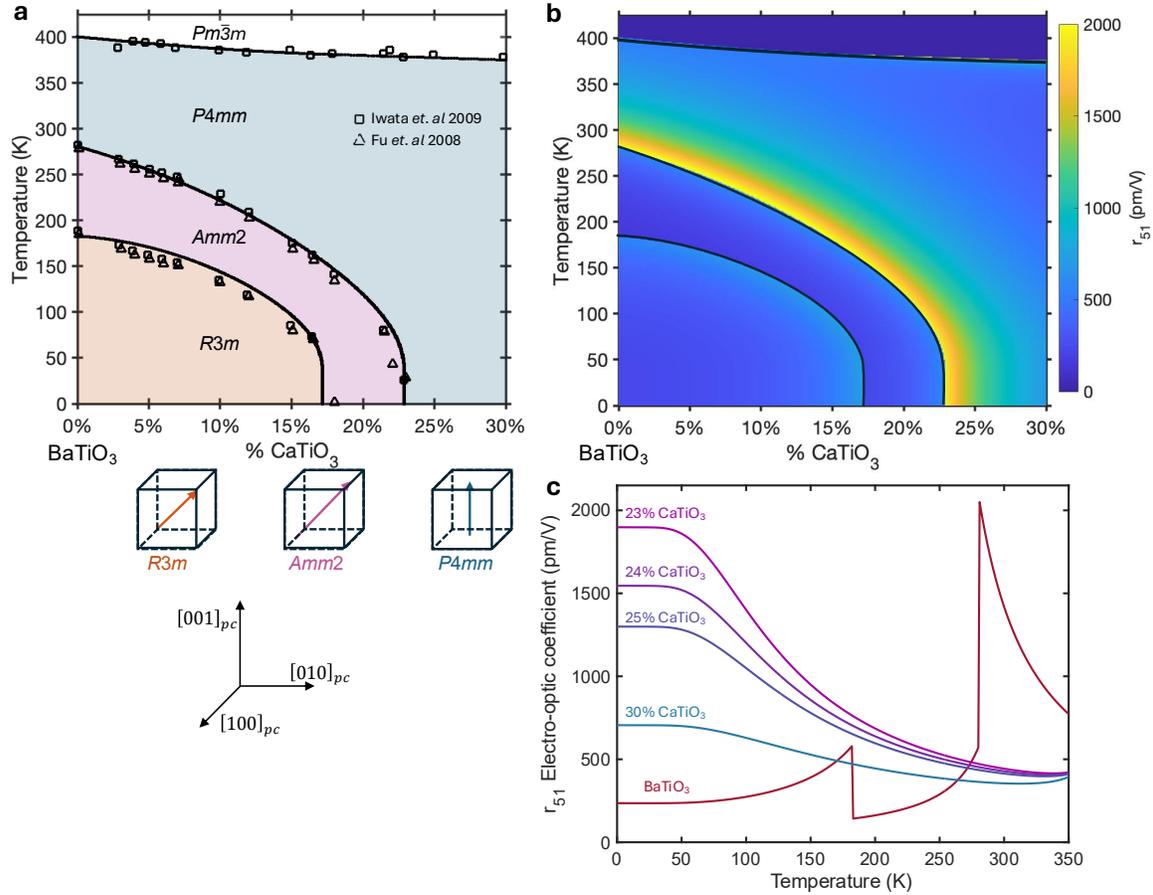

**Figure 4. Chemically engineered electro-optic response in $Ba_{1-x}Ca_xTiO_3$. a,** Temperature-composition phase diagram of $Ba_{1-x}Ca_xTiO_3$ showing the stability regions of the tetragonal, orthorhombic, and rhombohedral phases. **b,** Temperature and composition dependence of the calculated linear electro-optic $r_{51}$ for $Ba_{1-x}Ca_xTiO_3$ illustrating the relationship between the phase diagram and the electro-optic response. **c,** Calculated temperature dependence of the $r_{51}$ electro-optic coefficient under different $CaTiO_3$ concentrations.

**Figure 4a** presents the calculated temperature-composition diagram of $Ba_{1-x}Ca_xTiO_3$ under stress free conditions, compared with experimentally determined phase boundaries[33,34]. Increasing the $CaTiO_3$ fraction progressively stabilizes the tetragonal phase at the expense of the orthorhombic and rhombohedral shifting the phase boundaries to lower temperatures. Beyond 23% $CaTiO_3$ the phase diagram indicates that all ferroelectric–ferroelectric transitions are eliminated, leaving only the tetragonal phase (**Figure 4a**). The change in phase boundaries also change the temperature at which the $r_{51}$ electro-optic coefficient is calculated to be maximized (**Figure 4b**). As the concentration of CaTiO$_3$ is increased and the tetragonal to orthorhombic phase boundary is pushed to lower temperatures, the temperature where the $r_{51}$ electro-optic coefficient is calculated to be maximized is also pushed to lower temperatures. At 23% CaTiO$_3$ the tetragonal-to-orthorhombic phase boundary is tuned to 0 K which maximizes the calculated $r_{51}$ electro-optic coefficient directly within the quantum saturation regime. Beyond 23% CaTiO$_3$ the maximum calculated $r_{51}$ coefficient shifts to 0 K and the magnitude of the calculated $r_{51}$ coefficient decreases as the CaTiO$_3$ concentration is further increased (**Figure 4c**).

## Electro-Optic Performance in the Quantum Saturation Regime

**Figure 5** compares the temperature-dependent electro-optic coefficients of several representative materials, including strained BaTiO$_3$ on GdScO$_3$ (this work), Ba$_{0.77}$Ca$_{0.23}$TiO$_3$ (this work), bulk BaTiO$_3$[10,11], BaTiO$_3$ on Si[13], electrically biased SrTiO$_3$ and electrically biased SrTi(O$^{16}_{0.67}$O$^{18}_{0.33}$)$_3$[24]. In the high-temperature regime, large electro-optic coefficients are typically confined to a narrow temperature window, with small changes in temperature causing large changes in the electro-optic response, as seen in the behaviour of the $r_{51}$ coefficient of bulk BaTiO$_3$ near room temperature[10,11]. In contrast, in the quantum-saturated regime, a large electro-optic coefficient can be combined with a small thermal sensitivity so the response varies only weakly over a broad temperature range, as observed in the behaviour of BaTiO$_3$ on GdScO$_3$, Ba$_{0.77}$Ca$_{0.23}$TiO$_3$, electrically biased SrTiO$_3$ and electrically biased SrTi(O$^{16}_{0.67}$O$^{18}_{0.33}$)$_3$[24]. This illustrates that quantum saturation can overcome the usual trade-off between magnitude and thermal sensitivity. Notably, the cryogenic electro-optic response of BaTiO$_3$ on GdScO$_3$, is comparable to the peak electro-optic effect observed in bulk BaTiO$_3$ near room temperature. Moreover, the electro-optic coefficient is approximately 14 times larger than the cryogenic electro-optic coefficient measured in BaTiO$_3$ on Si and about 2.5 times larger than recent reports of electrically biased SrTi(O$^{16}_{0.67}$O$^{18}_{0.33}$)$_3$ [13,24].

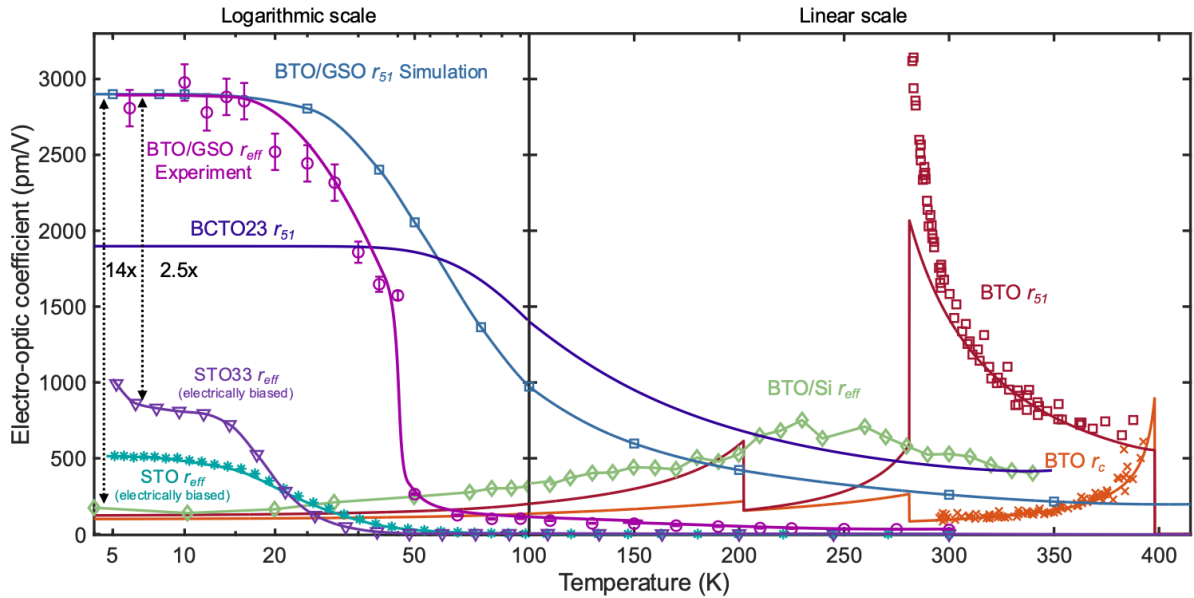

**Figure 5. Temperature-dependent electro-optic coefficients across different material platforms.** Temperature dependence of the linear electro-optic coefficient for bulk BaTiO$_3$ (BTO) (squares: experiment[10], solid line: theory[11]), Ba$_{0.77}$Ca$_{0.23}$TiO$_3$ (BCTO23) BaTiO$_3$ on Si[13] (BTO/Si), electrically biased SrTiO$_3$ (STO), and isotope-substituted SrTi(O$^{16}_{0.67}$O$^{18}_{0.33}$)$_3$ (STO33)[24] and BaTiO$_3$ on GdScO$_3$ (BTO/GSO) (circles: experiment, squares: theory).

## Discussion

In this work we demonstrate that the conventional trade-off between large electro-optic responses and thermal stability can be overcome by harnessing quantum fluctuations in ferroelectric materials. By tuning the ferroelectric phase boundaries to 0 K using strain or composition we engineer a cryogenic electro-optic response that is over an order of magnitude larger than strain-relaxed BaTiO$_3$. The combination of a large low-temperature electro-optic coefficient and a small thermal sensitivity demonstrates that strain/composition engineering and quantum saturation can be used to create high performance electro-optic materials for quantum technologies. We demonstrate that this design principle is not limited to epitaxially strained thin films and show that compositional tuning in Ba$_{1-x}$Ca$_x$TiO$_3$ offers a complementary route towards large cryogenic electro-optic properties which can be readily implemented in established commercial synthesis techniques, without the need of rare isotopes as required in isotope-substituted SrTi(O$^{16}_{0.67}$O$^{18}_{0.33}$)$_3$ [24].

Importantly, we find that merely stabilizing the tetragonal phase of BaTiO$_3$ is insufficient to sustain a large electro-optic coefficient at cryogenic temperatures, despite what may be inferred from the room temperature behaviour. While the tetragonal phase exhibits a large electro-optic response at room temperature, this enhancement primarily arises from its proximity to the tetragonal-to-orthorhombic phase boundary. Our findings underscore the importance of engineering phase competition to achieve large electro-optic effects. Although this current work takes advantage of quantum effects to stabilize phase competition over a broad temperature range, the underlying design principle promoting phase competition offers a broadly applicable framework for developing high-performance electro-optic materials across a wide range of temperatures.

Although currently these quantum effects are confined to the cryogenic regime, in principle these effects could be pushed to higher temperatures. In BaTiO$_3$ the saturation temperature ($T_s$) is set by the characteristic energy of the ferroelectric soft mode, but in principle, by changing the chemical composition or coupling to other lattice modes, this saturation temperature could be shifted to higher temperatures. This is analogous to the heat capacity of diamond near room temperature[35], where the high characteristic phonon energy pushes the onset of classical behaviour to high temperatures. By analogy, designing ferroelectrics with higher-frequency soft modes or engineering the coupling between other modes could raise the saturation temperature, opening the possibility of a room-temperature quantum-saturated electro-optic response[36].

The design strategy demonstrated here is not limited to BaTiO$_3$ or Ba$_{1-x}$Ca$_x$TiO$_3$ and extending the present analysis to other material systems could identify material platforms that can produce large cryogenic electro-optic responses. Moreover, these findings are not limited to the electro-optic effect and may be extended to other properties such as the piezoelectric or piezo-optic effect. In this sense, BaTiO$_3$ strained to GdScO$_3$ and Ba$_{1-x}$Ca$_x$TiO$_3$ serve as a model systems illustrating how phase competition and quantum fluctuations can be deliberately engineered to enable a high performance electro-optic response.

## Conclusion

We demonstrate that quantum fluctuations in ferroelectric materials can overcome the conventional trade-off between the magnitude and thermal sensitivity of the electro-optic response. By tuning the ferroelectric phase boundaries down to 0 K through either epitaxial strain or chemical composition, we engineer an electro-optic response that exceeds BaTiO$_3$ on Si by an order of magnitude while remaining insensitive to temperature below 25 K. Notably, Ba$_{1-x}$Ca$_x$TiO$_3$ enables electro-optic thin films with a large optical confinement which are compatible with established commercial synthesis methods. Although the quantum saturated regime explored here is confined to cryogenic temperatures, extending this framework to materials with higher-energy soft modes or tailored lattice couplings offer a compelling pathway to raise the saturation temperature. More broadly, this general design strategy is not limited to the electro-optic effect and may be an effective design strategy to piezoelectric and piezo-optic responses to support a broad range of cryogenic quantum technologies.

## Methods

### Phase-Field Simulations

Based on the thermodynamic theory of optical properties, we may formulate a phase-field model, where the thermodynamics of the ferroelectric system is described by its free energy functional,

$$F = \int \left[ f^L(T, P_i^L) + f^e(T, P_i^e) + f^{L-e}(P_i^L, P_i^e) + f^{grad}(\nabla_i P_j^L) + f^{elas}(P_i^L, P_i^e, \varepsilon_{ij}) + f^{elec}(P_i^L, P_i^e, E_i) \right] dx^3, \quad (7)$$

where $P_i^L(x_i, t)$ is the lattice polarization, $P_i^e(x_i, t)$ is the electrical polarization and $\sigma_{ij}(x, t)$ is the stress[11].

The respective evolution of the lattice polarization and electronic polarization is solved separately. The lattice polarization is computed following the relaxational approximation where we assume relaxational kinetics for the lattice polarization, and the mechanical displacement, electrical potential, and electrical polarization instantaneously reach equilibrium, leading to the time-dependent Ginzburg-Landau equation

$$\gamma_{ij}^L \frac{\partial P_j^L}{\partial t} = -\frac{\delta F}{\delta P_i^L}, \quad (8)$$

where $-\frac{\delta F}{\delta P_i^L}$ is the thermodynamic driving force for the temporal evolution of the lattice polarization. We use the local stress and lattice polarization distribution as inputs to the electronic polarization dynamic equation

$$\mu_{ij}^e \frac{\partial^2 P_j^e}{\partial t^2} + \gamma_{ij}^e \frac{\partial P_j^e}{\partial t} = -\frac{\delta F}{\delta P_i^e}, \tag{9}$$

where $-\frac{\delta F}{\delta P_i^e}$ is the thermodynamic driving force for the temporal evolution of the electronic polarization. To compute the local electronic dielectric susceptibility, we solve equation 3 assuming a small periodic electric field is applied, which yields an analytical solution

$$\widetilde{\chi_{ij}^{e,1}}(x_i, \omega) = \left[B_{ij}^e(x_i) - \epsilon_0\left(i\omega\gamma_{ij}^e + \omega^2\mu_{ij}^e\right)\right]^{-1}, \tag{10}$$

which is related to the curvature of the free-energy landscape with respect to the electronic polarization by, $B_{ij}^e(x_i) = \epsilon_0 \left(\frac{\partial^2 f}{\partial P_i^e \partial P_j^e}\right)$. Solving equation 14 (assuming $\lambda = 1550\ nm$) we may compute the local refractive indices from the local lattice polarization and stress distribution, thereby naturally including the electro-optic effect and potential piezo-optic effects. As the lattice polarization distribution evolves in response to an applied electric field, the corresponding change to the electronic dielectric susceptibility and the refractive index is computed.

Here $f^L(P_i^L)$ describes the intrinsic stability of the lattice polarization compared to the high symmetry phase ($m\bar{3}m$) as a Taylor expansion of the polarization about the high symmetry phase, this is equivalent to the Landau free energy density:

$$f^L(T, P_i^L) = f_o + a_{ij}(T)P_i^L P_j^L + a_{ijkl}P_i^L P_j^L P_k^L P_l^L + a_{ijklmn}P_i^L P_j^L P_k^L P_l^L P_m^L P_n^L + \tag{11}$$
$$a_{ijklmnop}P_i^L P_j^L P_k^L P_l^L P_m^L P_n^L P_o^L P_p^L,$$

where $a_{ij}$, $a_{ijkl}$, $a_{ijklmn}$ and $a_{ijklmnop}$ are the dielectric stiffness coefficients measured under constant stress conditions. For BaTiO$_3$, an 8$^{th}$ order expansion is used to describe the stability of the lattice polarization. The thermodynamic coefficients are adapted from ref. [25] to describe the effects of quantum fluctuations at cryogenic temperatures[23].

The intrinsic free energy density of the electronic polarization, in the absence of the lattice polarization, is described by

$$f^e(T, P_i^e) = \frac{1}{2\epsilon_0} B_{ij}^{e,ref}(T) P_i^e P_j^e, \tag{12}$$

where $B_{ij}^{ref}(T)$ is related to the refractive index of the equivalent high symmetry phase. The coupling energy density between the lattice and electronic polarization which determines the electro-optic effect is given by

$$f^{L-e}(P_i^L, P_i^e) = \frac{1}{2\epsilon_0} g_{ijkl}^{LL} P_l^L P_k^L P_i^e P_j^e, \tag{13}$$

where $g_{ijkl}^{LL}$ relates the lattice polarization to the refractive index.

The gradient energy density is represented by

$$f_{grad} = \frac{1}{2} G_{ijkl} \frac{\partial P_i^L}{\partial x_j} \frac{\partial P_k^L}{\partial x_l}, \tag{14}$$

where $G_{ijkl}$ is the gradient energy tensor where the non-zero coefficients are chosen to be $G_{11} = 0.6$, $G_{22} = -0.6$ and $G_{44} = 0.6$, and the units are normalized by $\alpha_1 l_o^2$, where $\alpha_1$ is the first Landau expansion coefficient and $l_o$ is chosen as 1 nm per grid point.

The elastic energy is given by

$$f^{elas}(P_i^L, P_i^e, \sigma_{ij}) = \frac{1}{2} C_{ijkl} (\varepsilon_{ij} - \varepsilon_{ij}^o)(\varepsilon_{kl} - \varepsilon_{kl}^o), \tag{15}$$

where $C_{ijkl}$ is the elastic stiffness, $\varepsilon_{ij}$ is the total strain, and $\varepsilon_{ij}^o = Q_{ijkl}\sigma_{ij}P_k^L P_l^L + \frac{1}{2\epsilon_0}\pi_{ijkl}\sigma_{ij}P_k^e P_l^e$ is the eigenstrain, where $Q_{ijkl}$ is the electrostrictive coefficient and $\pi_{ijkl}$ is the piezo-optic tensor for the high-symmetry phase. The total strain is solved for a thin film boundary condition assuming that the strain relaxes to its equilibrium value at each time step; for simplicity we ignore the contribution of the electronic polarization to the eigenstrain. Further details on solving the elasticity may be found in ref. [37].

The electrostatic energy is given by

$$f^{elec} = -E_i P_i^L - E_i P_i^e - \frac{1}{2}\epsilon_0 \kappa_{ij}^b E_i E_j, \tag{16}$$

where $\epsilon_0$ is the vacuum permittivity and $\kappa_{ij}^b$ is the background dielectric constant. For simplicity to compute evolution of the lattice polarization we only include the linear contribution to electronic polarization and add it to the background dielectric constant yielding $f^{elec} = -E_i P_i^L - \frac{1}{2}\epsilon_0 \kappa_{ij}^b E_i E_j$, where $\kappa_{ij}^b = 10$ and is isotropic. Here $\kappa_{ij}^b$ contains contributions from the electronic contribution and the vacuum and other hard modes. A table containing all coefficients used in the phase-field simulations is given in the supplementary material.

For the simulations, a system size of $128\Delta x_1$ $128 \Delta x_2 \times 36 \Delta x_3$ is used. There are 12 $\Delta x_3$ grid points for the substrate where the elastic constants are assumed to be the same as the ferroelectric film, and there are 4 $\Delta x_3$ grid points acting as an air layer above the film. The thickness of BaTiO3 is set to be 20 $\Delta x_3$. Periodic boundary conditions are used along the in-plane lateral directions, and natural boundary conditions are used along the film-substrate and film-air interfaces. The interface between the film and the substrate is assumed to be coherent, and hence the misfit strain is calculated using the equivalent cubic lattice parameters for BaTiO3 and the pseudocubic lattice parameters for GdScO3[38].

$$\varepsilon_{11} = \frac{a_{GdScO3}^{[110]_o} - a_{BaTiO3}^{eq}}{a_{BaTiO3}^{eq}}, \varepsilon_{22} = \frac{a_{GdScO3}^{[001]_o} - a_{BaTiO3}^{eq}}{a_{BaTiO3}^{eq}} \quad \varepsilon_{12} = \varepsilon_{21} = 0, \tag{17}$$

where the lattice parameters for BaTiO3 are $a_{BaTiO3}^{eq} = 4.008$ Å $(1 + 1.15 \times 10^{-5}(T - 300\text{ K}))$ and GdScO3 are $a_{GdScO3}^{[110]_o} = 3.970$ Å $(1 + 1.09 \times 10^{-5}(T - 300\text{ K}))$ and $a_{GdScO3}^{[001]_o} = 3.966$ Å $(1 + 1.09 \times 10^{-5}(T - 300\text{ K}))$.

The simulations begin with an initial condition of $P_i^L(x_i, t = 0) = [0.0\ 0.0\ 0.1\text{ C m}^{-2}] + \Delta P^{noise}(x_i, t = 0)$, where $|P^{noise}| = 0.1$ C m$^{-2}$, and relax for 20,000 time steps. To simulate the evolution of the polarization under an applied electric field and the corresponding electro-optic effect, a uniform electric field is applied along the $[100]_{pc}$ direction ranging from -70 kV cm$^{-1}$ to 70 kV cm$^{-1}$ which varies in time with a sinewave, with one period of the being equal to 65,000 time steps. The linear electro-optic effect is found by taking the average of the refractive index over the ferroelectric material as a function of an applied electric field and fitting to a cubic polynomial centred at zero. We note that the phase-field results reported here exhibit quantitative differences from our earlier work[19]. These discrepancies arise primarily from the use of a larger simulation domain and a slower electric-field ramp rate, choices made to more faithfully reproduce the experimental conditions. The lattice–polarization evolution was computed using the Mu-Pro phase-field simulation package[39].

## Thermodynamic Analysis

Using thermodynamic energy density function in equation 7, evaluate the ferroelectric and optical properties of Ba$_{1-x}$Ca$_x$TiO$_3$ assuming a single domain state under stress-free conditions. First, the equilibrium lattice and electronic polarizations are found by solving the following coupled equations

$$\left(\frac{\partial f}{\partial P_i^L}\right)_{T,\sigma,P^e} = 0, \left(\frac{\partial f}{\partial P_i^e}\right)_{T,\sigma,P^L} = 0. \tag{18}$$

Under zero applied electrical field, the electronic polarization by definition will be equal to zero[11].

The lattice dielectric susceptibility is then given by

$$\chi_{ij}^L = \frac{1}{\epsilon_0}\left(\frac{\partial P_i^L}{\partial E_j}\right)_{T,\sigma} = \frac{1}{\epsilon_0}\left(\frac{\partial^2 f}{\partial P_i^L \partial P_j^L}\right)_{T,\sigma}^{-1}. \tag{19}$$

To account for optical dispersion and loss, we can utilize the polarization equation of motion

$$\mu_{ij}^e \frac{\partial^2 P_j^e}{\partial t^2} + \gamma_{ij}^e \frac{\partial P_j^e}{\partial t} = \frac{\delta F}{\delta P_i^e} \tag{20}$$

Where $\mu_{ij}^e$ is the electronic polarization effective mass and $\gamma_{ij}^e$ is the electronic polarization damping coefficient. Solving equation 20 using a perturbation expansion yields

$$\widetilde{\chi_{ij}^e} = \tilde{n}^2 - \delta_{ij} = \left[B_{ij}^e - \epsilon_0(i\omega\gamma_{ij}^e + \omega^2\mu_{ij}^e)\right]^{-1}, \tag{21}$$

where $n$ is the refractive index and $B_{ij}^e = \frac{1}{\epsilon_0}\left(\frac{\partial^2 f}{\partial P_i^e \partial P_j^e}\right)_{T,\sigma}$ is the electronic dielectric stiffness. Which can be used to calculate the optical frequency dependent of the electro-optic effect via

$$r_{ijk}^L(\omega) = \left(\frac{\partial B_{ij}(\omega)}{\partial P_k^L}\right)_{T,\sigma}\left(\frac{\partial P_m^L}{\partial E_k}\right)_{T,\sigma} = f_{ijk}^L(\omega)\chi_{mk}^L \tag{22}$$

where

$$f_{ijk}^L = \left(\frac{\partial B_{ij}(\omega)}{\partial P_k^L}\right)_{T,\sigma} = \left(\frac{\partial B_{ij}(\omega)}{\partial B_{mn}^e}\right)_{T,\sigma}\left(\frac{\partial B_{mn}^e}{\partial P_k^e}\right)_{T,\sigma} = \left(\frac{\partial^3 f}{\partial P_i^e \partial P_j^e \partial P_k^L}\right)_{T,\sigma}$$

$$= [(B^e - \epsilon_0\omega^2\mu^e + \mathbb{I})^{-1}]_{mi}[(B^e - \epsilon_0\omega^2\mu^e + \mathbb{I})^{-1}]_{nj}[2g_{mnkl}^{LL}P_l^L + g_{mnkl}^{Le}P_l^e]. \tag{23}$$

In all calculations, we find the frequency $\omega$ assuming $\lambda = 1550\ nm$. The thermodynamic coefficients for $Ba_{1-x}Ca_xTiO_3$ are given in table S2 and were fit to experimental measurements of the spontaneous polarization, phase transition temperatures and dielectric constants[33,34,40,41]. For simplicity, we ignore the composition dependence of the quadratic polar-optic tensor, and reference electronic dielectric stiffness and use the values obtained for $BaTiO_3$.

### Growth of $BaTiO_3$ on $GdScO_3$ Thin Films

The sample was grown in a Veeco GEN10 MBE system equipped with an Epiray GmbH THERMALAS laser substrate heater, which is a 1 kW $CO_2$ laser with a wavelength of 10.6 μm that irradiates the backside of the substrate over a circular area with a diameter of approximately 14 mm. Barium (Sigma-Aldrich, 99.99% purity) was supplied using a conventional differentially pumped effusion cell and titanium using a Veeco Ti-Ball source. The sources were simply co-deposited. The barium flux was approximately $5 \times 10^{13}$ atoms/(cm$^2$ × s) and the titanium flux was approximately $1 \times 10^{13}$ atoms/(cm$^2$ × s), i.e., a Ba:Ti flux ratio of 5:1. The barium flux was determined by a quartz crystal microbalance, and the titanium flux by x-ray reflectivity of a calibration film of $BaTiO_3$ grown on an $SrTiO_3$ (001) substrate at $T_{sub}$ = 1200 °C, with an oxidant pressure of $1 \times 10^{-6}$ Torr of $O_2$ + 10% $O_3$ and a Ba:Ti ratio of 5:1. The $BaTiO_3$ film was grown on a $GdScO_3$ (110)$_O$ substrate (CrysTec GmbH), to a thickness of approximately 36 nm.

Ozone was used as the oxidant at a background pressure of $1 \times 10^{-6}$ Torr of $O_2$ + 10% $O_3$. The film was cooled in the same oxidant and pressure in which it was grown to $T_{sub}$ < 200 °C before the oxidant was turned off. The substrate temperature, measured by a pyrometer operating at 7.5 μm on the backside of the substrate during growth, was 1160 °C. Due to the high substrate temperature, the vapor pressure of barium-containing species over $BaTiO_3$, chiefly BaO, is significantly higher than that of titanium-containing species over $BaTiO_3$, so excess barium will desorb from the surface leaving behind a single-phase $BaTiO_3$ film within an adsorption-controlled growth window. As $GdScO_3$ absorbs well at 10.6 μm, the backside of the substrate was not coated.

X-ray diffraction (XRD) and reciprocal space mapping (RSM) analysis was done with a Panalytical Empyrean x-ray diffractometer using Cu $K_{\alpha 1}$ radiation. Atomic force microscopy (AFM) was done using an Asylum Cypher ES Environmental AFM. RHEED images were taken during growth using a Staib electron source operating at 14 kV.

### Polarization-Based Electro-Optic Measurement

Electro-optic measurements were performed using a homebuilt PSCA setup with a 1550 nm HP 81689A continuous-wave laser source and a Nirvana 2017 balanced detector. For absolute intensity measurements, a mechanical chopper was placed in the laser path and modulated at 1.54 kHz for detection with a Stanford SR830 lock-in amplifier. For measuring the modulated transmission intensity, an Agilent 33220A signal generator amplified by a Trek 610C high voltage amplifier was used to apply a 555 Hz sine wave to the sample with the lock-in amplifier reference signal changed to the function generator reference. Samples were cooled to liquid He temperatures using a Janis 30 continuous flow cryostat system.

The polarization of the probe is set to lie between the ordinary and extraordinary refractive indices. An analyzer placed at the end of the setup is oriented orthogonal to the initial probe polarization state, resulting in no transmission of light through the setup when the sample and compensator are not included. With the sample introduced, the transmission becomes nonzero on account of the birefringence-induced ellipticity. A quarter-wave compensator is then added immediately after the sample to cancel out the native material birefringence with no voltage applied. This returns the transmission of the setup to zero until a voltage is applied to modulate the sample refractive indices and reintroduce ellipticity to the probe.

### Second-Harmonic Generation Polarimetry

Second-harmonic generation polarimetry experiments were performed using a fundamental wavelength of 800 nm generated by a Spectra-Physics Solstice Ace amplified Ti:Sapphire laser with a 1 kHz repetition rate and 100 fs pulse width. Light was focused onto the sample using a 10 cm focal length lens, producing a 50 μm diameter spot size as determined by knife edge measurements. Samples were cooled to liquid He temperatures using a Janis 30 continuous flow cryostat system. A bare $GdScO_3$ $(110)_O$ substrate of the same type produces no appreciable SHG signal under the same incident power, and thus the SHG signal measured is interpreted as originating entirely from the strained $BaTiO_3$ film.

## Acknowledgements

The work was primarily supported by the National Science Foundation under award number DMR-2522897. A.R. acknowledges the support of the National Science Foundation Graduate Research Fellowship Program under Grant No. DGE1255832. The phase-field simulations in this work were performed using Bridges-2 at the Pittsburgh Supercomputing Center through allocation MAT230041 from the ACCESS program, which is supported by National Science Foundation grants #2138259, #2138286, #2138307, #2137603 and #2138296. The $BaTiO_3$ thin films were synthesized at the Platform for the Accelerated Realization, Analysis, and Discovery of Interface Materials (PARADIM), which is supported by the National Science Foundation (NSF) under Cooperative Agreement No. DMR-2039380. D.S. and D.G.S. acknowledge support from the NSF through PARADIM under Cooperative Agreement No. DMR-2039380.

## Data Availability

The data that supports the findings of this study are available within the article. Datasets used for the generation of the figures are available at https://doi.org/10.5281/zenodo.17842835. Additional information relating to the film growth and structural characterization is available at https://doi.org/10.34863/k6vg-fr77. Any additional data connected to the study are available from the corresponding author upon reasonable request.

## Competing interests

The authors have a provisional patent filed on this work. Long-Qing Chen has a financial interest in MuPRO, LLC, a company which licenses and markets the software package used in this research.

Supplementary Information

## S1: Parameters in the thermodynamic free-energy function for BaTiO3

For BaTiO3, we use the cubic phase ($m\bar{3}m$) as our high symmetry reference state and employ an 8th-order landau expansion describe the relative stability of the lattice polarization compared to the cubic reference state. The coefficients used for this paper are adjusted from Li et al.[25] and Ross et al.[11] and to include the effect of cryogenic fluctuations and are given in Table S1.

**Table S1.** Coefficients in the thermodynamic free energy function and equation of motion for BaTiO3

| | | | |
|---|---|---|---|
| $g^{LL}_{1111}$ | $18.5 \times 10^{-2}\,(\mathrm{m^4/C^2})$ | $p_{1111}$ | 0.5328 (Unitless) |
| $g^{LL}_{1122}$ | $2.5 \times 10^{-2}\,(\mathrm{m^4/C^2})$ | $p_{1122}$ | 0.1584 (Unitless) |
| $g^{LL}_{1212}$ | $12.85 \times 10^{-2}\,(\mathrm{m^4/C^2})$ | $p_{1212}$ | $-0.432$ (Unitless) |
| $\mu_e$ | $35.5 \times 10^{-23}\,\left(\frac{\mathrm{Kg\,m^4}}{\mathrm{m\,C^2}}\right)$ | $\gamma_e$ | $3 \times 10^{-9}\,\left(\frac{\mathrm{Kg\,m^4}}{\mathrm{ms\,C^2}}\right)$ |
| $a_{11}$ | $a_0\, T_s\left(\coth\left(\frac{T_s}{T}\right) - \coth\left(\frac{T_s}{T_c}\right)\right)$ | $B^{e,ref}_{ij}(T_0)$ | 0.2356 (Unitless) |
| $T_c$ | 388 K | $T_0$ | 398 K |
| $T_s$ | 54 K | $a_0$ | $4.124 \times 10^{5}\,\left(\frac{\mathrm{J}}{\mathrm{m^3}}\,\frac{\mathrm{m^4}}{\mathrm{K\,C^2}}\right)$ |
| $a_{1111}$ | $-2.097 \times 10^{8}\,\left(\frac{\mathrm{J}}{\mathrm{m^3}}\,\frac{\mathrm{m^8}}{\mathrm{C^4}}\right)$ | $\alpha_{11}$ | $2.657 \times 10^{-5}\,\left(\frac{1}{\mathrm{K}}\right)$ |
| $a_{1122}$ | $7.974 \times 10^{8}\,\left(\frac{\mathrm{J}}{\mathrm{m^3}}\,\frac{\mathrm{m^8}}{\mathrm{C^4}}\right)$ | $a_{11111111}$ | $3.863 \times 10^{10}\,\left(\frac{\mathrm{J}}{\mathrm{m^3}}\,\frac{\mathrm{m^{16}}}{\mathrm{C^8}}\right)$ |
| $a_{111111}$ | $1.294 \times 10^{9}\,\left(\frac{\mathrm{J}}{\mathrm{m^3}}\,\frac{\mathrm{m^{12}}}{\mathrm{C^6}}\right)$ | $a_{11111122}$ | $2.529 \times 10^{10}\,\left(\frac{\mathrm{J}}{\mathrm{m^3}}\,\frac{\mathrm{m^{16}}}{\mathrm{C^8}}\right)$ |
| $a_{111122}$ | $-1.95 \times 10^{9}\,\left(\frac{\mathrm{J}}{\mathrm{m^3}}\,\frac{\mathrm{m^{12}}}{\mathrm{C^6}}\right)$ | $a_{11112222}$ | $1.637 \times 10^{10}\,\left(\frac{\mathrm{J}}{\mathrm{m^3}}\,\frac{\mathrm{m^{16}}}{\mathrm{C^8}}\right)$ |
| $a_{112233}$ | $-2.509 \times 10^{9}\,\left(\frac{\mathrm{J}}{\mathrm{m^3}}\,\frac{\mathrm{m^{12}}}{\mathrm{C^6}}\right)$ | $a_{11112222}$ | $1.637 \times 10^{10}\,\left(\frac{\mathrm{J}}{\mathrm{m^3}}\,\frac{\mathrm{m^{16}}}{\mathrm{C^8}}\right)$ |
| $C_{11}$ | $1.78 \times 10^{11}$ Pa | $C_{44}$ | $1.22 \times 10^{11}$ Pa |
| $C_{12}$ | $0.964 \times 10^{11}$ Pa | | |

## S2: Parameters in the thermodynamic free-energy function for Ba$_{1-x}$Ca$_x$TiO$_3$

For Ba$_{1-x}$Ca$_x$TiO$_3$, we use the cubic phase ($m\bar{3}m$) as our high symmetry reference state and employ an 8$^{th}$-order landau expansion describe the relative stability of the lattice polarization compared to the cubic reference state. The thermodynamic coefficients for Ba$_{1-x}$Ca$_x$TiO$_3$ are given in table S2 and were fit to experimental measurements of the spontaneous polarization, phase transition temperatures and dielectric constants[33,34,40,41]. For simplicity, we ignore the composition dependence of the quadratic polar-optic tensor, and reference electronic dielectric stiffness and use the values obtained for BaTiO$_3$. The coefficients used for Ba$_{1-x}$Ca$_x$TiO$_3$ are given in table S2.

**Table S2.** Coefficients in the thermodynamic free energy function and equation of motion for Ba$_{1-x}$Ca$_x$TiO$_3$

| | | | |
|---|---|---|---|
| $g^{LL}_{1111}$ | $18.5 \times 10^{-2} \, (m^4/C^2)$ | $p_{1111}$ | 0.5328 (Unitless) |
| $g^{LL}_{1122}$ | $2.5 \times 10^{-2} \, (m^4/C^2)$ | $p_{1122}$ | 0.1584 (Unitless) |
| $g^{LL}_{1212}$ | $12.85 \times 10^{-2} \, (m^4/C^2)$ | $p_{1212}$ | $-0.432$ (Unitless) |
| $\mu_e$ | $35.5 \times 10^{-23} \left(\frac{Kg \, m^4}{m \, C^2}\right)$ | $\gamma_e$ | $3 \times 10^{-9} \left(\frac{Kg \, m^4}{ms \, C^2}\right)$ |
| $a_{11}$ | $a_0 T_s \left(\coth\left(\frac{T_s}{T}\right) - \coth\left(\frac{T_s}{T_c}\right)\right)$ | $B^{e,ref}_{ij}(T_0)$ | 0.2356 (Unitless) |
| $T_c$ | $390(1-x) + 340x$ K | $T_0$ | 398 K |
| $T_s$ | 140 K | $a_0$ | $4.124 \times 10^5 \left(\frac{J}{m^3} \frac{m^4}{K \, C^2}\right)$ |
| $a_{1111}$ | $-2.097 \times 10^8 (1 - 6.0x) \left(\frac{J}{m^3} \frac{m^8}{C^4}\right)$ | $\alpha_{11}$ | $2.657 \times 10^{-5} \left(\frac{1}{K}\right)$ |
| $a_{1122}$ | $7.974 \times 10^8 (1 + 3.6x + 7.0x^2) \left(\frac{J}{m^3} \frac{m^8}{C^4}\right)$ | $a_{11111111}$ | $3.863 \times 10^{10} \left(\frac{J}{m^3} \frac{m^{16}}{C^8}\right)$ |
| $a_{111111}$ | $1.294 \times 10^9 \left(\frac{J}{m^3} \frac{m^{12}}{C^6}\right)$ | $a_{11111122}$ | $2.529 \times 10^{10} \left(\frac{J}{m^3} \frac{m^{16}}{C^8}\right)$ |
| $a_{111122}$ | $-1.75 \times 10^9 (1 + 5.0x) \left(\frac{J}{m^3} \frac{m^{12}}{C^6}\right)$ | $a_{11112222}$ | $1.637 \times 10^{10} \left(\frac{J}{m^3} \frac{m^{16}}{C^8}\right)$ |
| $a_{112233}$ | $-2.509 \times 10^9 (1 + 32x) \left(\frac{J}{m^3} \frac{m^{12}}{C^6}\right)$ | $a_{11112233}$ | $1.367 \times 10^{10} \left(\frac{J}{m^3} \frac{m^{16}}{C^8}\right)$ |